\documentclass{ifacconf}
\usepackage{graphicx}
  \graphicspath{{./pics/}}
  \DeclareGraphicsExtensions{.pdf}
\usepackage{natbib}
\usepackage{xspace}
\usepackage{tikz}
  \usetikzlibrary{arrows,automata,shapes,backgrounds}
  \usetikzlibrary{decorations.pathmorphing}
  \usetikzlibrary{calc}
\usepackage{amssymb,amsmath}
\usepackage{newtxtext,newtxmath}

\usepackage[textwidth=1.9cm,color=green!10,textsize=footnotesize]{todonotes}

\allowdisplaybreaks

\newcommand{\eps}{\varepsilon}
\DeclareMathOperator{\supN}{\rm supN}
\DeclareMathOperator{\supn}{\rm supN}

\DeclareMathOperator{\supRO}{\rm supRO}

\newcommand{\complclass}[1]{{\sc #1}\xspace}
\newcommand{\PSpace}{\complclass{PSpace}}

\begin{document}
\begin{frontmatter}

\title{Conditions for Hierarchical Supervisory Control under Partial Observation\thanksref{footnoteinfo}} 

\thanks[footnoteinfo]{Partially supported by RVO 67985840 and by the GA\v{C}R grant GC19-06175J.}

\author[First]{Jan Komenda} 
\author[Second,First]{Tom{\' a}{\v s} Masopust} 

\address[First]{Institute of Mathematics of the Czech Academy of Sciences, Brno, Czechia}
\address[Second]{Faculty of Science, Palacky University, Olomouc, Czechia \\ 
	(e-mails: komenda{@}ipm.cz, masopust{@}math.cas.cz)}

\begin{abstract}
  The fundamental problem in hierarchical supervisory control under partial observation is to find conditions preserving observability between the original (low-level) and the abstracted (high-level) plants. Two conditions for observable specifications were identified in the literature -- {\em observation consistency\/} (OC) and {\em local observation consistency} (LOC). However, the decidability of OC and LOC were left open. We show that both OC and LOC are decidable for regular systems. We further show that these conditions do not guarantee that supremal (normal or relatively observable) sublanguages computed on the low level and on the high level always coincide. To solve the issue, we suggest a new condition -- {\em modified observation consistency} -- and show that under this condition, the supremal normal sublanguages are preserved between the levels, while the supremal relatively observable high-level sublanguage is at least as good as the supremal relatively observable low-level sublanguage, i.e., the high-level solution may be even better than the low-level solution.
\end{abstract}
 

\end{frontmatter}

\onecolumn

\section*{Errata}

Several results of this paper are not correct as stated. This section contains errata based on the revised and extended version of this conference paper.\footnote{Komenda, J., Masopust, T., Hierarchical Supervisory Control under Partial Observation: Normality, 2023,
https://doi.org/10.48550/arXiv.2203.01444} No changes are made in the following sections.

\medskip

 (1) The statement of~Theorem~\ref{thm5}, claiming that the verification of (modified) observation consistency is \PSpace-complete for NFAs, is unproved. In particular, the membership in \PSpace is not shown. The problem is \PSpace-hard, but it is open whether the verification of (M)OC is decidable; in particular, it is open whether the problem is in \PSpace.

\medskip
 (2) The statement of Theorem~\ref{supN} is not precise. The correct statement is:
  \setcounter{thm}{10}
  \begin{thm}
    For a nonblocking DFA $G$, let $L=L(G)$ and $L_m=L_m(G)$. If $L$ is MOC with respect to $Q$, $P$, and $P_{hi}$, then for every high-level specification $K\subseteq Q(L_m)$,
    \[
      \supN\bigl(K\| L_m,L,P\bigr)
        = \supN(K,Q(L),P_{hi}) \parallel L_m
    \]
    whenever $\supN(K,Q(L),P_{hi})$ and $L_m$ are nonconflicting.
  \end{thm}
  \setcounter{thm}{0}

\medskip

  (3) As stated, Theorem~\ref{2n} is incorrect.
    \setcounter{thm}{14}
  \begin{thm}
    Assume that each shared event is high level and observable, i.e., $\Sigma_{s}\subseteq \Sigma_{hi} \cap \Sigma_{o}$. If, for $i=1,\ldots,n$, $L_i$ is MOC wrt $Q_i$, $P_{loc}^{i}$, and $P_{loc|hi}^{i}$, then $\|_{i=1}^n L_i$ is MOC wrt $Q$, $P$, and $P_{hi}$.
  \end{thm}
  A counterexample:
  Let $L_1 = \overline{\{h_1o_1x\}}$ and $L_2= \overline{\{o_2h_2x\}}$ be two languages with observable events $\Sigma_o=\{o_1,o_2,x\}$ and high-level events $\Sigma_{hi}=\{h_1,h_2,x\}$; that is, $x$ is the only shared event, which is both observable and high-level.
  Both languages satisfy MOC.
  However, for $s=h_1o_1o_2h_2x$ and $t=h_2h_1x$ satisfying $P_{hi}(Q(s)) = P_{hi}(h_2h_1x) = x = P_{hi}(h_1h_2x) = P_{hi}(t)$, there is no $s'\in L_1 \| L_2 = h_1o_1x \| o_2h_2x$ such that $P(s') = o_1o_2x$ and $Q(s') = h_2h_1x$, because the only string containing $x$ with the order of observable events $o_1o_2$ is $h_1o_1o_2h_2x$, but it does not have the required order of the local high-level events $h_1$ and $h_2$.

  \setcounter{thm}{0}

\twocolumn

\section{Introduction}
  Organizing systems into hierarchical structures is a common engineering practice used in manufacturing, robotics, or artificial intelligence to overcome the combinatorial state explosion problem. 
  Hierarchical supervisory control of discrete-event systems (DES) was introduced by \cite{ZhongW1990} as a two-level vertical decomposition of the system. The low-level plant modeling the system behavior is restricted by a high-level specification, and the aim is to synthesize a nonblocking and optimal supervisor based on the high-level abstraction of the plant in such a way that it can be used for a low-level implementation. They identified a sufficient condition to achieve the goal. \cite{ZhongW1990b} extended the framework to hierarchical coordination control and developed an abstract hierarchical supervisory control theory. \cite{WongW96a} applied the theory to the Brandin-Wonham framework of timed DES.
  \cite{KS} extended hierarchical supervisory control to decentralized systems, and \cite{SB11} found weaker sufficient conditions for maximal permissiveness of high-level supervisors with complete observations. Recently, \cite{BaierM15} generalized hierarchical supervisory control to the B\"uchi framework, where the plant and the specification are represented by $\omega$-languages.
 
  Motivated by abstractions of hybrid systems to DES, \cite{HubbardC02} developed a hierarchical control theory for DES based on state aggregation, and \cite{TorricoC2002} investigated a hierarchical control approach where the low level is in the Ramadge-Wonham framework and the high level is obtained by state aggregation. Here, the high-level events are subsets of low-level events, and advanced control structures are used to synthesize a controller. Furthermore, \cite{CunhaC07} proposed hierarchical supervisory control for DES where the low level is in the Ramadge-Wonham framework and the high level is represented by systems with flexible marking, in order to simplify the modeling of the high level.
  \cite{NgoS14,NgoS18} investigated hierarchical control for Moore automata and for timed DES, and \cite{SakakibaraU2018} considered concurrent DES modeled by Mealy automata.
  
  \cite{fekri2009} first considered hierarchical supervisory control of partially observed DES. They used Moore automata models and defined controllable and observable events based on vocalization. Hence, they need a specific definition of the low-level supervisor. Furthermore, their approach is monolithic, while ours allows distributed synthesis using the standard synchronous composition of the plant with the supervisor.

  In this paper, we adapt the classical hierarchical supervisory control of DES in the Ramadge-Wonham framework, where the systems are modeled as DFAs and the abstraction is modeled as a natural projection, i.e., the behavior of the high-level plant is the projection of the behavior of the low-level plant to the high-level alphabet. The problem is then as follows. Given a low-level plant $G$ over an alphabet $\Sigma$ modeling the system behavior and a high-level specification language $K$ over a high-level alphabet $\Sigma_{hi}\subseteq \Sigma$. The low-level plant $G$ is abstracted to the high-level plant $G_{hi}$ describing the high-level behavior. The aim is to synthesize a nonblocking and optimal supervisor $S_{hi}$ on the high level in such a way that it can be used for a construction of a low-level supervisor $S$ that is nonblocking and optimal wrt the specification $K\| L_m(G)$. 

  To achieve the goal for fully observed DES, important concepts have been developed in the literature, including the {\em observer property\/} of \cite{WW96}, {\em output control consistency\/} (OCC) of \cite{ZhongW1990}, and {\em local control consistency\/} (LCC) of \cite{SB11}. These concepts are sufficient for the high-level synthesis of a nonblocking and optimal supervisor to have a low-level implementation.
  
  However, the conditions are not sufficient for partially observed DES. The sufficient condition of \cite{KM10} requires that all observable events must be high-level events, which is a very restrictive assumption.
  Therefore, \cite{cdc-ecc2011} investigated weaker and less restrictive conditions, and introduced two concepts -- {\em local observation consistency\/} (LOC) and {\em observation consistency\/} (OC). The latter ensures a certain consistency between observations on the high level and the low level, and the former is an extension of the observer property to partial observation. The paper shows that, for observable specifications, projections that satisfy OC, LOC, LCC, and that are observers are suitable for the nonblocking least restrictive hierarchical supervisory control under partial observation. The fundamental question whether the properties of OC and LOC are decidable is left open. 

  In this paper, we first show that checking OC and LOC properties is decidable for systems with regular behaviors and that the problems are actually \PSpace-complete (Theorems~\ref{thm5} and~\ref{thm7}). 

  Then we show that OC and LOC are not sufficient to preserve optimality for non-observable specifications. These are specifications, for which a suitable supremal sublanguage (normal or relatively observable) needs to be computed. We show that OC and LOC do not guarantee that the supremal normal (relatively observable) low-level sublanguage coincides with the composition of the plant and the supremal normal (relatively observable) high-level sublanguage (Example~\ref{counterexample1}). 
  
  For normality, we suggest a condition of {\em modified observation consistency} (MOC) and show that it preserves optimality, i.e., the supremal normal sublanguages are preserved between the levels (Definition~\ref{defMOC} and Theorem~\ref{supN}). Then we discuss two special cases often considered in the literature: (i) the case where all observable events are also high-level events, and (ii) the case where all high-level events are also observable. Our new results generalize the previously known results.
  
  For relative observability, we show that MOC ensures that the high-level solution is at least as good as the low-level solution (Theorem~\ref{supRO}). In particular, the low-level implementation of the high-level solution may be better than what we can obtain directly on the low level (Example~\ref{counterexample3}). This observation makes relative observability an interesting and suitable notion for hierarchical supervisory control.

  Finally, the newly suggested condition of MOC is stronger than OC of \cite{cdc-ecc2011} as shown in Lemma~\ref{MOCstrongerOC}. Moreover, similarly as OC, the MOC condition is structural only wrt the plant. We discuss the complexity of MOC in Theorem~\ref{MOCcomplexity}, and show that it is compositional in Theorem~\ref{2n}.

  All the missing proofs can be found in the appendix.

\section{Preliminaries and Definitions}
  We assume that the reader is familiar with the basics of supervisory control, see \cite{CL08}.
  For a set $A$, $|A|$ denotes the cardinality of $A$. For an alphabet (finite nonempty set) $\Sigma$, $\Sigma^*$ denotes the set of all finite strings over $\Sigma$; the empty string is denoted by $\eps$. 
  The alphabet $\Sigma$ is partitioned into {\em controllable events} $\Sigma_c$ and {\em uncontrollable events} $\Sigma_u=\Sigma\setminus\Sigma_c$ as well as into {\em observable events} $\Sigma_o$ and {\em unobservable events} $\Sigma_{uo}=\Sigma\setminus\Sigma_o$. A language is a subset of $\Sigma^*$. For a language $L\subseteq \Sigma^*$, the prefix closure $\overline{L}=\{w\in \Sigma^* \mid wv\in L\}$; $L$ is prefix-closed if $L=\overline{L}$.
  
  A {\em (natural) projection\/} $R\colon \Sigma^* \to \Gamma^*$, where $\Gamma \subseteq \Sigma$ are alphabets, is a homomorphism for concatenation defined so that $R(a) = \eps$ for $a\in \Sigma\setminus \Gamma$, and $R(a) = a$ for $a\in \Gamma$. The action of $R$ on $w\in\Sigma^*$ is to remove all events from $w$ that are not in $\Gamma$. The inverse image of $w\in\Gamma^*$ under $R$ is the set $R^{-1}(w)=\{s\in \Sigma^* \mid R(s) = w\}$. These definitions can naturally be extended to languages. 
  
  A {\em nondeterministic finite automaton\/} (NFA) is a quintuple $G = (Q,\Sigma,\delta,I,F)$, where $Q$ is a finite set of states, $\Sigma$ is an input alphabet, $I\subseteq Q$ is a set of initial states, $F\subseteq Q$ is a set of marked states, and $\delta \colon Q\times\Sigma \to 2^Q$ is the transition function that can be extended to the domain $2^Q\times \Sigma^*$ in the usual way. 
  The automaton $G$ is {\em deterministic\/} (DFA) if $|I|=1$, and $|\delta(q,a)|=1$ for every state $q \in Q$ and every event $a \in \Sigma$. 
  The language {\em generated\/} by $G$ is the set $L(G) = \{w\in \Sigma^* \mid \delta(q_0,w)\in Q\}$, and the language {\em marked\/} by $G$ is the set $L_m(G) = \{w\in \Sigma^* \mid \delta(q_0,w)\in F\}$.
  By definition, $L_m(G)\subseteq L(G)$, and $L(G)$ is prefix-closed. If $\overline{L_m(G)} = L(G)$, then $G$ is {\em nonblocking}.
  
  Let $L_1\subseteq \Sigma_1^*$, $L_2\subseteq \Sigma_2^*$ be languages. The {\em parallel composition of $L_1$ and $L_2$\/} is the language $L_1\| L_2 = P_1^{-1}(L_1) \cap P_2^{-1}(L_2)$, where $P_i\colon (\Sigma_1\cup\Sigma_2)^* \to \Sigma_i^*$ is a projection, for $i=1,2$; see \cite{CL08} for a definition for automata. For two DFAs $G_1$ and $G_2$, $L(G_1 \| G_2) = L(G_1) \| L(G_2)$. Languages $L_1$ and $L_2$ are {\em synchronously nonconflicting\/} if $\overline{L_{1}\|L_{2}} = \overline{L_{1}} \| \overline{L_{2}}$.

  Let $G$ be a DFA over an alphabet $\Sigma$. A language $K\subseteq L_m(G)$ is {\em controllable} wrt $L(G)$ and the set of uncontrollable events $\Sigma_u$ if $\overline{K}\Sigma_u\cap L(G)\subseteq \overline{K}$;
  $K$ is {\em observable\/} wrt $L(G)$, the set of observable events $\Sigma_o$ with $P\colon \Sigma^*\to \Sigma_o^*$ being the corresponding projection, and the set of controllable events $\Sigma_c$ if for all $s,s'\in L(G)$ with $P(s)=P(s')$ and for every $e\in \Sigma_c$, if $se \in \overline{K}$, $s'e \in L(G)$, and $s' \in \overline{K}$, then $s'e \in \overline{K}$. Algorithms to verify controllability and observability can be found in \cite{CL08}. 
  
  It is known that there is no supremal observable sublanguage. Therefore, stronger properties, such as normality of \cite{LinWon88} or relative observability of \cite{CaiZW15}, are used for specifications that are not observable.
  Language $K\subseteq L_m(G)$ is {\em normal\/} wrt $L(G)$ and the projection $P\colon\Sigma^*\to\Sigma_o^*$ if $\overline{K} = P^{-1}[P(\overline{K})]\cap L(G)$.
  Relative observability has recently been introduced by \cite{CaiZW15} and further studied by \cite{AlvesCB17} as a condition weaker than normality and stronger than observability. Let $K \subseteq C \subseteq L_m(G)$ be languages. Language $K$ is {\em relatively observable\/} wrt $C$, $G$, and $P\colon \Sigma^*\to\Sigma_o^*$ (or simply {\em $C$-observable}) if for all strings $s, s'\in \Sigma^*$ with $P(s) = P(s')$ and for every $e \in \Sigma$, whenever $se \in \overline{K}$, $s'e\in L(G)$, and $s' \in \overline{C}$, then $s'e \in \overline{K}$. For $C=K$, the definition coincides with observability.

  A {\em decision problem\/} is a yes-no question. A decision problem is {\em decidable\/} if there exists an algorithm that solves the problem. Complexity theory classifies decidable problems to classes based on the time or space an algorithm needs to solve the problem. The complexity class we consider in this paper is \PSpace, denoting all problems solvable by a deterministic polynomial-space algorithm. A decision problem is \PSpace-complete if the problem belongs to \PSpace ({\em membership}) and every problem from \PSpace can be reduced to the problem by a polynomial-time algorithm ({\em hardness}). 
  It is unknown whether \PSpace-complete problems can be solved in polynomial time.

\section{Principles of Hierarchical Control}
\begin{figure}[h]
    \centering
    \begin{tikzpicture}[>=stealth',auto,sloped,bend angle=45,baseline,->,shorten >=1pt,thick,node distance=4cm]
      \node[] (2) {$\Sigma^*$};
      \node[] (4) [below right=0.3cm and 2cm of 2] {$\Sigma_{o}^*$};
      \node[] (6) [above right=0.3cm and 2cm of 2] {$\Sigma_{hi}^*$};
      \node[] (8) [below right=0.3cm and 2cm of 6] {$(\Sigma_o \cap \Sigma_{hi})^*$};
      
      \path 
        (2) edge[above] node{$P$} (4)
        (4) edge[above] node{$Q_{o}$} (8)
        (6) edge[above] node{$P_{hi}$} (8)
        (2) edge[above] node{$Q$} (6)
      ;
    \end{tikzpicture}
    \caption{Commutative diagram of abstractions and projections.}
    \label{projections}
  \end{figure}
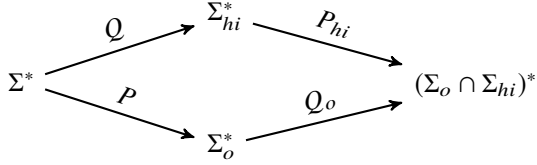

  In the sequel, we use the following notation for projections and abstractions, see the commutative diagram in Fig.~\ref{projections}. Let $\Sigma$ be the low-level alphabet, $\Sigma_{hi}\subseteq \Sigma$ the high-level alphabet, and $\Sigma_o\subseteq \Sigma$ the set of observable events. Let $P\colon \Sigma^* \to \Sigma^*_o$ be the projection corresponding to system's partial observation, $Q\colon \Sigma^* \to \Sigma_{hi}^*$ the projection corresponding to the high-level abstraction, and
  $P_{hi}\colon \Sigma_{hi}^{*} \to (\Sigma_{hi}\cap \Sigma_o)^*$ and
  $Q_o\colon \Sigma_o^* \to (\Sigma_{hi}\cap \Sigma_o)^*$ the corresponding observations and abstractions.

  We now state the hierarchical supervisory control problem for partially observed DES.
  
  \begin{prob}\label{prob1}
    Let $G$ be a low-level plant over an alphabet $\Sigma$, and let $K$ be a high-level specification over an alphabet $\Sigma_{hi} \subseteq \Sigma$. The abstracted high-level plant $G_{hi}$ is defined over the alphabet $\Sigma_{hi}$ so that $L(G_{hi}) = Q(L(G))$ and $L_m(G_{hi}) = Q(L_m(G))$. The aim of hierarchical supervisory control is to determine, based on the high-level plant $G_{hi}$ and the specification $K$, without using the low-level plant $G$, a nonblocking low-level supervisor $S$ such that $L_m(S/G) = K \| L_m(G)$.
    \hfill$\diamond$
  \end{prob}
    
  \cite{cdc-ecc2011} identified sufficient conditions (observation consistency and local observation consistency) on the low-level plant $G$ for which observability of $K\| L_m(G)$ wrt $G$ is equivalent to observability of $K$ wrt the high-level plant $G_{hi}$.
  
  A prefix-closed language $L \subseteq \Sigma^*$ is {\em observation consistent} (OC) wrt projections $Q$, $P$, and $P_{hi}$ if for all strings $t,t' \in Q(L)$ such that $P_{hi}(t) = P_{hi}(t')$, there are $s,s' \in L$ such that $Q(s) = t$, $Q(s') = t'$, and $P(s) = P(s')$. Intuitively, any two strings of the high-level plant with the same observation have corresponding strings with the same observation in the low-level plant.

  A prefix-closed language $L \subseteq \Sigma^*$ is {\em locally observation consistent} (LOC) wrt projections $Q$ and $P$ and the set of controllable events $\Sigma_c$ if for all strings $s,s'\in L$ and all events $e\in \Sigma_c \cap \Sigma_{hi}$ such that $Q(s)e, Q(s')e \in Q(L)$ and $P(s) = P(s')$, there exist low-level strings $u,u'\in (\Sigma\setminus \Sigma_{hi})^*$ such that $P(u) = P(u')$ and $sue, s'u'e\in L$.
  Intuitively, continuing two observationally equivalent high-level strings by the same controllable event, the corresponding low-level observationally equivalent strings can be continued by this same event in the original plant in the future (after possible empty low-level strings with the same observations). LOC can be seen as a specialization of the observer property and LCC for partially observed DES.
  
  Besides observability, Problem~\ref{prob1} further requires the preservation of controllability between the levels. It has been previously achieved by the conditions of {\em $L_m(G)$-observer} of \cite{WW96} and {\em output control consistency} of \cite{ZhongW1990}, or its weaker variant, \emph{local control consistency} of \cite{SB11}.
  Formally, projection $Q\colon \Sigma^* \to \Sigma_{hi}^*$ is an {\em $L_m(G)$-observer} for a nonblocking plant $G$ over $\Sigma$ if for all strings $t\in Q(L_m(G))$ and $s\in \overline{L(G)}$, if $Q(s)$ is a prefix of $t$, then there exists $u\in \Sigma^*$ such that $su\in L_m(G)$ and $Q(su)=t$.
  We say that $Q$ is {\em locally control consistent\/} (LCC) for a string $s\in L(G)$ if for all $e \in \Sigma_{hi} \cap \Sigma_u$ such that $Q(s)e \in L(G_{hi})$, either there is no $u\in (\Sigma \setminus \Sigma_{hi})^*$ such that $sue \in L(G)$ or there is $u\in (\Sigma_u\setminus \Sigma_{hi})^*$ such that $sue\in L(G)$. We call $Q$ LCC for a language $M\subseteq L(G)$ if $Q$ is LCC for every $s\in M$. 
  
  Notice that the conditions are structural and hold for any specification once the plant is fixed. The following result formulates a solution to Problem~\ref{prob1}.
    
  \begin{thm}[\cite{cdc-ecc2011}]\label{coro}
    Let $G$ be a nonblocking DFA over $\Sigma$, and let $K\subseteq Q(L_m(G))$ be a (high-level) specification. Let $Q$ be LCC for $L(G)$ and $\Sigma_u$, and an $L_m(G)$-observer. Let $L(G)$ be OC wrt $Q$, $P$, and $P_{hi}$, and LOC wrt $Q$, $P$, and $\Sigma_c$. Then $K$ is controllable wrt $Q(L(G))$ and $\Sigma_u \cap  \Sigma_{hi}$, and observable wrt $Q(L(G))$, $\Sigma_o \cap \Sigma_{hi}$, and $\Sigma_c \cap \Sigma_{hi}$ if and only if $K\|L_m(G)$ is controllable wrt $L(G)$ and $\Sigma_u$, and observable wrt $L(G)$, $\Sigma_o$, and $\Sigma_c$.
    \hfill$\qed$
  \end{thm}
  
  Theorem~\ref{coro} allows to verify the existence of a supervisor realizing a high-level specification $K$ for a given system $G$, under the aforementioned properties, based on the abstraction $G_{hi}$. Namely, if there is a nonblocking supervisor $S_{hi}$ such that $L_m(S_{hi}/G_{hi}) = K$, then there is a nonblocking supervisor $S$ such that $L_m(S/G) = K \| L_m(G)$. In particular, a DFA realization $G_K$ of $K$ such that $L_m(G_K) = K$ can be used to implement the supervisor in the form $G_K \| G$.

  Considering only observability, the following results hold.
  \begin{thm}[\cite{cdc-ecc2011}]\label{main}
    Let $G$ be a nonblocking DFA over $\Sigma$, and let $K\subseteq Q(L_m(G))$ be a specification. Assume that $L(G)$ is OC wrt $Q$, $P$, and $P_{hi}$, that $K$ and $L_m(G)$ are synchronously nonconflicting, and that $L(G)$ is LOC wrt $Q$, $P$, and $\Sigma_{c}$. Then $K$ is observable wrt $Q(L(G))$, $\Sigma_{hi} \cap \Sigma_o$, and $\Sigma_{hi} \cap \Sigma_c$ if and only if $K\| L_m(G)$ is observable wrt $L(G)$, $\Sigma_o$, and $\Sigma_c$.
    \hfill$\qed$
  \end{thm}

  If all controllable events are observable, observability is equivalent to normality, and OC is sufficient to preserve observability.
  
  \begin{cor}[\cite{cdc-ecc2011}]\label{cormain}
    Let $G$ be a nonblocking DFA, and let $K\subseteq Q(L_m(G))$ be a specification. If $L(G)$ is OC wrt $Q$, $P$, and $P_{hi}$, and $K$ and $L_m(G)$ are synchronously nonconflicting, then $K$ is normal wrt $Q(L(G))$ and $P_{hi}$ if and only if $K\| L_m(G)$ is normal wrt $L(G)$ and $P$.
    \hfill$\qed$
  \end{cor}

  We now show that a result similar to Theorem~\ref{main} does not hold for relative observability without additional assumptions; namely, if $K$ is $C$-observable, then $K\| L_m(G)$ is not necessarily $C\| L(G)$-observable. 
  Let $K=\{\eps,a\}$, $C=\{\eps,a,au\}$ over $\Sigma_{hi}=\{a,u\}$, and $L(G)=\{\eps,a,ae,au,aue\}$ over $\Sigma=\{a,u,e\}$ be prefix-closed languages, and hence synchronously nonconflicting. Let $\Sigma_o=\{a,e\}$. It can be verified that $L(G)$ is OC and LOC, and that $K$ is $C$-observable wrt $Q(L(G))=C$, and hence observable. However, $K\| L(G)$ is not $C\|L(G)$-observable, since $ae\in K\| L(G)$, $au\in C\| L(G)$, and $aue\in L(G)$, but $aue\notin K\|L(G)$ (but  $K\| L(G)$ is observable by Theorem~\ref{main}).

\section{Verification of Observation Consistency}
  In this section, we show that the verification of OC is \PSpace-complete, and hence decidable, for systems modeled by finite automata. The same problem for LOC is treated in the next section.
  
  \begin{thm}\label{thm5}
    Verifying OC for systems modeled by NFAs is \PSpace-complete.
  \end{thm}
  \begin{pf}
    To prove membership in \PSpace, we generalize the parallel composition to a set of synchronizing events. Let $\Sigma$ be an alphabet, and let $L_1,L_2\subseteq \Sigma^*$ be languages of NFAs $G_1=(Q_1,\Sigma,\delta_1,I_1,F_1)$ and $G_2=(Q_2,\Sigma,\delta_2,I_2,F_2)$, respectively. Let $\Sigma'\subseteq\Sigma$ be a set of synchronizing events. The parallel composition of $L_1$ and $L_2$ synchronized on the events of $\Sigma'$ is denoted by $L_1 \|_{\Sigma'} L_2$ and defined as the language of the NFA
    \[
      G_1 \|_{\Sigma'} G_2 = (Q_1\times Q_2, (\Sigma\cup\{\eps\})\times(\Sigma\cup\{\eps\}), \delta, I_1\times I_2, F_1\times F_2)\,,
    \]
    where the alphabet is a set of pairs based on the synchronization of events in $\Sigma'$. There are two categories of pairs to construct, corresponding to (a) events in $\Sigma'$, and (b) events in $\Sigma\setminus\Sigma'$. For every $a\in \Sigma'$, we have the pair $(a,a)$, and for every $a\in \Sigma\setminus \Sigma'$, we have two pairs $(a,\eps)$ and $(\eps,a)$. The transition function $\delta\colon (Q_1\times Q_2) \times ((\Sigma\cup\{\eps\})\times (\Sigma\cup\{\eps\})) \to Q_1\times Q_2$ is defined on these event pairs as follows:
    \begin{itemize}
      \item for $a\in\Sigma'$, 
        $\delta((p,q),(a,a)) = \delta_1(p,a) \times \delta_2(q,a)$;
      \item for $a\in\Sigma\setminus\Sigma'$, 
        $\delta((p,q),(a,\eps)) = \delta_1(p,a)\times \{q\}$ and 
        $\delta((p,q),(\eps,a)) = \{p\} \times \delta_2(q,a)$;
      \item undefined otherwise.
    \end{itemize}
    
    For simplicity, a sequence of event pairs, $(a_1,\eps)(a_2,a_2)(\eps,a_3)$, is written as a pair of the concatenated components $(a_1a_2,a_2a_3)$. Then we can say that the language consists of pairs of strings of the form $(w,w')$, where $w$ and $w'$ coincide on the letters of $\Sigma'$, that is, $P'(w)=P'(w')$ for the projection $P'\colon \Sigma^* \to \Sigma'^*$. 
    
    Let $L \subseteq \Sigma^*$ be a prefix-closed language, and let $\Sigma_o$ and $\Sigma_{hi}$ be the respective observation and high-level alphabets. We show that $L$ is OC wrt $Q$, $P$, and $P_{hi}$ if and only if
    \[
      Q(L)\parallel_{\Sigma_{hi}\cap \Sigma_o} Q(L) \subseteq Q\left(L\parallel_{\Sigma_o} L\right)\,,
    \]
    where, for an event $(a,b)$, $Q(a,b)=(Q(a),Q(b))$.     Membership in \PSpace then follows, since we can express $Q(L)$, as well as $Q(L\parallel_{\Sigma_o} L)$, as NFAs, and the inclusion of two NFAs can be verified in \PSpace, see \cite{ClementeM19}.
    
    The intuition behind the equivalence is to couple all strings $t,t'\in Q(L)$ with the same high-level observations, which are exactly the pairs $(t,t')\in Q(L)\parallel_{\Sigma_{hi}\cap \Sigma_o} Q(L)$, and to verify that for every such pair there are strings $s,s' \in L$ with the same observations, which are exactly the pairs $(s,s')\in L\parallel_{\Sigma_o} L$, that are abstracted to the pair $(t,t')$, that is, they satisfy $(Q(s),Q(s'))=(t,t')$. 

    The rest of the proof can be found in the appendix.
  \hfill$\qed$\end{pf}
  
  By a slight modification of the proof, it can be shown that the problem is not easier for DFAs, that is, it remains \PSpace-hard even for DFA models. We leave this proof for the full version.

\section{Verification of Local Observation Consistency}
  In this section, we study decidability and complexity of LOC. As in the case of OC, the problem is not easier for DFA models. The proof is again left for the full version. 
  A proof sketch of the following theorem can be found in the appendix.
  
  \begin{thm}\label{thm7}
    Verification of LOC for systems modeled by NFAs is \PSpace-complete.
  \end{thm}

\section{Preservation of Supremality}
  Problem~\ref{prob1} requires that the specification language $K$ is achievable by the supervisor, i.e., $K$ is observable. However, this is not always the case. If $K$ is not observable, a common approach is to find a suitable sublanguage of $K$ that is observable. Since there is no supremal observable sublanguage, the supremal normal sublanguage or the supremal relatively observable sublanguage is computed instead. The problem is now formulated as follows.
  
  \begin{prob}\label{prob2}
    Given a low-level plant $G$ over $\Sigma$ and a high-level specification $K$ over $\Sigma_{hi} \subseteq \Sigma$. The abstracted high-level plant $G_{hi}$ over $\Sigma_{hi}$ is defined so that $L(G_{hi}) = Q(L(G))$ and $L_m(G_{hi}) = Q(L_m(G))$. The aim is to determine a maximally permissive nonblocking supervisor $S$ such that $L_m(S/G) \subseteq K \| L_m(G)$ using the abstraction $G_{hi}$. That is, if a maximally permissive nonblocking supervisor $S_{hi}$ exists for the abstracted plant such that $L_m(S_{hi}/G_{hi}) \subseteq K$, then a maximally permissive nonblocking supervisor $S$ exists such that $L_m(S/G) \subseteq K \| L_m(G)$. 
  \end{prob}
  
  Compared to Corollary~\ref{cormain} saying that under the OC condition the specification $K$ is normal if and only if $K\|L_m(G)$ is normal, the following example shows that OC is not sufficient to preserver normality (relative observability) if the supremal normal (relatively observable) sublanguage of the specification $K$ is a strict sublanguage of $K$. The problem is that it is not true that every supremal normal (relatively observable) sublanguage of $K\| L_m(G)$ is of the form $X\| L_m(G)$ for some convenient language $X\subseteq K$, and hence there may be no $X$ that would be the supremal normal sublanguage of $K$.

  Before stating the example, we introduce the following notation. For a prefix-closed language $L$ and a specification $K\subseteq L$, we write $\supN(K,L)$ (resp. $\supRO(K,L)$) to denote the supremal normal (resp. the supremal relatively observable) sublanguage of $K$ wrt $L$ and the corresponding set of observable events.
  
  \begin{exmp}\label{counterexample1}
    Let $\Sigma=\{a,b,c\}$ with $\Sigma_o=\{a,c\}$ and $\Sigma_{hi}=\{b,c\}$, and let
    $
      L=\{\eps,a,b,c,ba,ac,bac\} \text{ and } K=\{\eps,b,c\}\subseteq Q(L)=\{\eps,b,c,bc\}\,.
    $
    To show that $L$ is OC, notice that $P_{hi}(\eps)=\eps=P_{hi}(b)$ and $P_{hi}(c)=c=P_{hi}(bc)$, and hence we have two cases: (i) $t=\eps$ and $t'=b$, and (ii) $t=c$ and $t'=bc$. Case (i) is trivial because we can choose $s=t=\eps$ and $s'=t'=b$, which clearly satisfies OC. For case (ii), we choose $s=ac$ and $s'=bac$. Then, $Q(s)=c=t$, $Q(s')=bc=t'$, and $P(s)=ac=P(s')$. Thus, $L$ is OC.
    
    To compute the supremal normal sublanguages, we use the formula of \cite{brandt} stating that $\supN(B,M)=B-P^{-1}P(M-B)\Sigma^*$, for prefix-closed languages $B\subseteq M \subseteq \Sigma^*$, and we obtain the following:
      $K\| L = a^*ba^* \cup a^*ca^* \cup a^* \cap L =\{\eps,a,b,c,ba,ac\}$,
      $L-K\| L=\{bac\}$, and
      $P^{-1}P(bac)=P^{-1}(ac)=b^*ab^*cb^*$.
    This gives that
    $
      c\in \supn(K\| L,L) = K\| L -P^{-1}P(L-K\|L)\Sigma^* = \{\eps,a,b,c,ba\}\,. 
    $
    On the other hand,
     $Q(L)-K=\{\eps,b,c,bc\}-\{\eps,b,c\}=\{bc\}$,
     $P_{hi}(bc)=c$, and $P_{hi}^{-1}(c)=b^*cb^*$,
    which gives that 
    $
      c\notin \supn(K,Q(L)) \| L = Q^{-1}(K-P_{hi}^{-1}P_{hi}(Q(L)-K)\Sigma_{hi}^*)\cap L = Q^{-1}(\{\eps,b\}) \cap L = \{\eps,a,b,ba\}
    $ showing that OC is not a sufficient condition to preserve supremal normal sublanguages.
    
    Inspecting further the example, the reader may verify that the computed supremal normal sublanguages coincide with the supremal relatively observable sublanguages for the choice of $C=K$. Therefore, the example also illustrates that OC is neither a sufficient condition to preserve supremal relatively observable sublanguages.
    \hfill$\Diamond$
  \end{exmp}

  To preserver the properties for supremal sublanguages, we modify the condition of OC by fixing one of the components.
  
  \begin{defn}\label{defMOC}
    A prefix-closed language $L \subseteq \Sigma^*$ is {\em modified observation consistent} (MOC) wrt projections $Q$, $P$, and $P_{hi}$ if for every $s\in L$ and every $t' \in Q(L)$ such that $P_{hi}(Q(s)) = P_{hi}(t')$, there exists $s' \in L$ such that $P(s) = P(s')$ and $Q(s') = t'$.
  \end{defn}
  
  MOC is a stronger property than OC. Indeed, if $L$ is MOC, then for any $t,t'\in Q(L)$ with $P_{hi}(t)=P_{hi}(t')$, we have that $t=Q(s)$ for some $s\in L$, and hence there exists $s'\in L$ such that $P(s)=P(s')$ and $Q(s')=t'$, which shows that $L$ is OC. This proves the following observation.
  \begin{lem}\label{MOCstrongerOC}
    MOC implies OC. \hfill$\qed$
  \end{lem}

\subsection{Normality}
  We now show that MOC guarantees the preservation of normality for supremal sublanguages.
  
  \begin{thm}\label{supN}
    Let $G$ be a nonblocking DFA, and let $K\subseteq Q(L_m(G))$ be a specification. If $L(G)$ is MOC wrt $Q$, $P$, and $P_{hi}$, and $K$ and $L_m(G)$ are synchronously nonconflicting, then 
    \[
      \supN\bigl(K\| L_m(G),L(G)\bigr)
        = \supN(K,Q(L(G))) \parallel L_m(G)\,.
    \]
  \end{thm}
  \begin{pf}
    ($\supseteq$): Since $\supN(K,Q(L(G)))$ is normal wrt $Q(L(G))$ and $P_{hi}$, Corollary~\ref{cormain} implies that $\supN(K,Q(L(G))) \parallel L_m(G)$ is normal wrt $L(G)$ and $P$. The implication that normality of $K$ implies normality of $K\|L_m(G)$ in Corollary~\ref{cormain} holds without any assumptions. Therefore, $\supN\bigl(K,Q(L(G))\bigr) \parallel L_m(G) \subseteq \supN\bigl(K\| L_m(G),L(G)\bigr)$.
    
    ($\subseteq$): Let $S\subseteq K\| L_m(G)$ be normal wrt $L(G)$ and $P$, that is, $\overline{S} = P^{-1}P(\overline{S}) \cap L(G)$. Then, $Q(S)\subseteq K\cap Q(L_m(G)) = K$. We show that $Q(S)$ is normal wrt $Q(L(G))$ and $P_{hi}$, i.e., that $Q(\overline{S}) = P_{hi}^{-1} P_{hi} (Q(\overline{S})) \cap Q(L(G))$. To do this, let $s\in \overline{S}$ and $t'\in Q(L(G))$ be such that $P_{hi}(Q(s))=P_{hi}(t')$, that is, $t'\in P_{hi}^{-1}P_{hi}(Q(\overline{S})) \cap Q(L(G))$. We show that $t'\in Q(\overline{S})$. By MOC, there exists $s'\in L(G)$ such that $Q(s')=t'$ and $P(s)=P(s')$, i.e., $s'\in P^{-1}P(s)\cap L(G) \subseteq P^{-1}P(\overline{S}) \cap L(G) = \overline{S}$, and hence $t'=Q(s')\in Q(\overline{S})$, which shows normality of $Q(S)$.
  \hfill$\qed$\end{pf}

  Two special cases are often considered in the literature:
    (i) $\Sigma_o\subseteq \Sigma_{hi}$, and
    (ii) $\Sigma_{hi}\subseteq \Sigma_{o}$.
  We show that both imply MOC, and hence OC. 
  Consequently, Theorem~\ref{supN} strengthens the result of \cite{KM10} showing that for any prefix-closed languages $L\subseteq \Sigma^*$ and $K\subseteq Q(L)$, if $\Sigma_o\subseteq \Sigma_{hi}$, then $\supn(K,Q(L))\parallel L = \supn(K\| L,L)$. 
  
  First, assume that $\Sigma_o\subseteq \Sigma_{hi}$. Then $P=P_{hi}Q$, since $Q_o$ is an identity. Let $s\in L$ and $t'\in Q(L)$ be such that $P_{hi}(Q(s))=P_{hi}(t')$. Consider any $s'\in L$ with $Q(s')=t'$; such $s'$ exists because $t'\in Q(L)$. Then, $P(s) = P_{hi}(Q(s)) = P_{hi}(t') = P_{hi}(Q(s')) = P(s')$, which was to be shown.

  Second, assume that $\Sigma_{hi}\subseteq \Sigma_{o}$. Then, $P_{hi}$ is an identity, and hence for any $s\in L$ and $t'\in Q(L)$ satisfying $P_{hi}(Q(s)) = P_{hi}(t')$, we have $Q(s) = P_{hi}(Q(s)) = P_{hi}(t') = t'$, i.e., we can chose $s'=s$ in the definition of MOC.

\subsection{Relative Observability}
  We now show that an analogy of Theorem~\ref{supN} does not hold for relative observability. In particular, the inclusion
  \[
    \supRO\bigl(K\| L_m(G),L(G)\bigr)
      \supseteq \supRO(K,Q(L(G))) \parallel L_m(G)
  \]
  does {\em not\/} hold in general as shown in the following example.

  \begin{exmp}\label{counterexample3}
    Let the low-level plant and the high-level specification be defined by automata in Fig.~\ref{fig3}.
    \begin{figure}
      \centering
      \includegraphics[scale=.78]{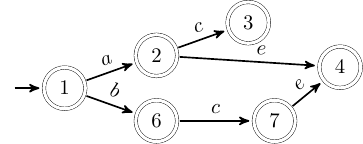}
      \includegraphics[scale=.78]{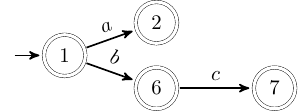}
      \caption{Plant $G$ and a specification $K$}
      \label{fig3}
    \end{figure}
    Let  $\Sigma_{hi}=\{a,b,c\}$ and $\Sigma_o=\{e\}$. Then $\supRO(K,Q(L(G)))$ is shown in Fig.~\ref{fig3a} as well as $\supRO(K,Q(L(G)))\parallel L_m(G)$. There, the reader can also see the supremal relatively observable sublanguage of $K\|L_m(G)$ wrt $K\|L_m(G)$, $L(G)$, and $P$, which obviously does not include $\supRO(K,Q(L(G)))\parallel L_m(G)$. \hfill$\diamond$ 
  \end{exmp}

  \begin{figure} 
    \centering
    \includegraphics[scale=.78]{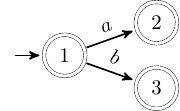}
    \includegraphics[scale=.78]{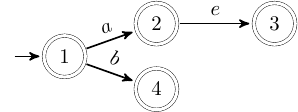}
    \includegraphics[scale=.78]{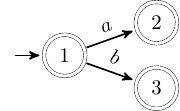}
    \caption{Languages $\supRO(K,Q(L(G)))$, $\supRO(K,Q(L(G)))\parallel L_m(G)$, and $\supRO(K\|L_m(G),L(G))$, respectively}
    \label{fig3a}
  \end{figure}
  
  By Theorem~\ref{main}, $\supRO(K,Q(L(G))) \parallel L_m(G)$ is always observable. It is thus an interesting question under which conditions the opposite inclusion holds. In other words, under which conditions is the low-level implementation of the high-level supervisor at least as good as the low-level supervisor? We now show that MOC is such a condition. 
  
  \begin{thm}\label{supRO}
    Let $G$ be a nonblocking DFA over $\Sigma$ and $K\subseteq Q(L_m(G))$ a specification. If $L(G)$ is MOC wrt $Q$, $P$, and $\Sigma_c$, and $K$ and $L_m(G)$ are synchronously nonconflicting, then 
    \[
      \supRO(K\|L_m(G),L(G)) \subseteq \supRO(K,Q(L(G))) \parallel L_m(G)\,.
    \]
  \end{thm}
  \begin{pf}
    Let $S=\supRO(K\|L_m(G),L(G))$. Since $S\subseteq K\| L_m(G)$, $Q(S)\subseteq K\cap Q(L_m(G)) = K$. We now show that $Q(S)$ is relatively observable wrt $K$, $Q(L(G))$, and $P_{hi}$. To this end, let $t,t'\in\Sigma_{hi}^*$ be such that $P_{hi}(t)=P_{hi}(t')$, and let $e\in\Sigma_{hi}$ be such that $te \in Q(\overline{S})$, $t'\in \overline{K}$, and $t'e \in Q(L(G))$. We have to show that $t'e\in Q(\overline{S})$. To this aim, let $se\in\overline S$ be such that $Q(se)=te$. Since $t'e \in Q(L(G))$ and $P_{hi}(Q(se)) = P_{hi}(t'e)$, MOC implies that there is $w' \in L(G)$ such that $Q(w') = t'e$ and $P(se) = P(w')$. Then $w'=s'e$ for some $s'\in L(G)$. Since $Q(w')=t'e$, we have that $Q(s')=t'$ and $P(s)=P(s')$. From $t' \in \overline K$ and the synchronous nonconflictingness of $K$ and $L_m(G)$, we conclude that $s' \in \overline{K}\|L(G)=\overline{K\|L_m(G)}$. Altogether, $P(s) = P(s')$, $se\in \overline{S}$, $s' \in \overline{K\|L_m(G)}$, and $s'e \in L(G)$. Then, relative observability of $S$ wrt $K\|L_m(G)$, $L(G)$, and $P$ implies that $s'e \in \overline S$. Hence, $t'e = Q(s'e) \in Q(\overline{S})$.
  \hfill$\qed$\end{pf}

  Notice that the plant in Example~\ref{counterexample3} does not satisfy MOC, and hence MOC is not a necessary condition in Theorem~\ref{supRO}.

  A proof of the following result can be found in the appendix.
  \begin{thm}\label{MOCcomplexity}
    Verifying MOC for NFAs is \PSpace-complete.
  \end{thm}

  Similarly as for OC, the verification of MOC is not easier for DFA models. We provide a proof of \PSpace-hardness for DFAs in the full version.

\section{Modularity}
  Let $G=G_1\|\cdots\| G_n$ be a modular DES. For simplicity, we write $L_i$ to denote $L(G_i)$ and $L=L(G)=L_1 \| \cdots \| L_n$. Similarly for $L_{m,i}=L_m(G_i)$ and $L_m=L_m(G)$.

\begin{figure}
    \centering
    \includegraphics[scale=1]{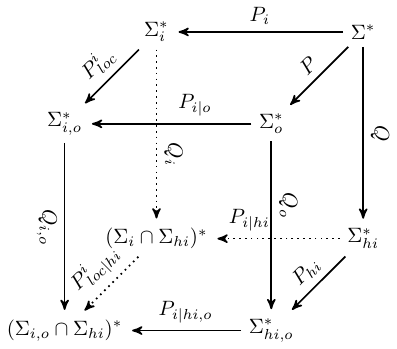}
    \caption{Our notation for the used projections}
    \label{fig1}
  \end{figure}

  In addition to the high-level alphabet $\Sigma_{hi}$ and the set of observable events $\Sigma_o$, we have the local alphabets $\Sigma_i$, $i=1,\dots,n$. The intersection of the alphabets is denoted by adding two corresponding subscripts, e.g., $\Sigma_{i,o}=\Sigma_i \cap \Sigma_o$ denotes the locally observable events of $\Sigma_i$, and $\Sigma_{hi,o}=\Sigma_{hi} \cap \Sigma_o$ denotes the high-level observable events. The various projections are denoted as shown in Fig.~\ref{fig1}.
  
  We further assume that the high-level alphabet contains all shared events, i.e.,
  $\Sigma_{s}\subseteq \Sigma_{hi}$,
  where $\Sigma_s=\cup_{i\not =j} (\Sigma_i\cap \Sigma_j)$ is the set of all events shared by two or more components. In addition, we assume that the modular components agree on the controllability and observability status of the shared events, which is a standard assumption in hierarchical decentralized control.

  We now show that if all the local languages satisfy MOC, the their parallel composition also satisfies MOC. 

  \begin{thm}\label{2n}
    Assume that each shared event is high level and observable, i.e., $\Sigma_{s}\subseteq \Sigma_{hi} \cap \Sigma_{o}$. If, for $i=1,\ldots,n$, $L_i$ is MOC wrt $Q_i$, $P_{loc}^{i}$, and $P_{loc|hi}^{i}$, then $\|_{i=1}^n L_i$ is MOC wrt $Q$, $P$, and $P_{hi}$.
  \end{thm}

\section{Conclusion}
  We have completed the missing results in hierarchical supervisory control under partial observation. 
  The regular behavior of the systems is essential for decidability of OC, MOC, and LOC. In the full version, we show that if slightly more expressive one-turn deterministic pushdown systems are used, the properties are undecidable. Deterministic pushdown systes have been discussed in supervisory control in the context of controllability and synthesis as a generalization of system models for which the synthesis is still possible.

\bibliography{biblio}

\appendix
\section{Proofs}
\subsection{\PSpace-hardness proof of Theorem~\ref{thm5}}
    We first show that if $L$ is OC, then the inclusion holds. To this end, assume that $(t,t')\in Q(L) \parallel_{\Sigma_{hi}\cap\Sigma_o} Q(L)$. By the definition of $\parallel_{\Sigma_{hi}\cap\Sigma_o}$, $t,t' \in Q(L)$ and $t,t'$ coincide on the letters of $\Sigma_{hi}\cap\Sigma_o$, i.e., $P_{hi}(t) = P_{hi}(t')$. Since $L$ is OC, there are $s,s' \in L$ such that $Q(s) = t$, $Q(s') = t'$, and $P(s) = P(s')$. However, $P(s)=P(s')$ implies that $(s,s')\in L\parallel_{\Sigma_o} L$, and $Q(s)=t$ and $Q(s')=t'$ imply that $(Q(s),Q(s'))=(t,t')$, which shows the inclusion.
    
    On the other hand, assume that the inclusion holds. We show that $L$ is OC. To this end, assume that $t,t'\in Q(L)$ are such that $P_{hi}(t) = P_{hi}(t')$. By the definition of $\parallel_{\Sigma_{hi}\cap\Sigma_o}$, we obtain that $(t,t')\in Q(L)\parallel_{\Sigma_{hi}\cap\Sigma_o} Q(L)$. Since the inclusion holds, we have $(t,t')\in Q(L \parallel_{\Sigma_o} L)$, which means that there is a pair $(s,s')\in L\parallel_{\Sigma_o} L$ such that $(Q(s),Q(s'))=(t,t')$. Since $(s,s')\in L\parallel_{\Sigma_o} L$, strings $s$ and $s'$ belong to $L$ and coincide on the letters from $\Sigma_o$, i.e., $P(s)=P(s')$, which was to be shown.

    To show \PSpace-hardness, we reduce the problem of deciding universality for NFAs with all states marked, see \cite{KaoRS09}. Such NFAs recognize exactly prefix-closed languages. The problem asks, given an NFA $A$ over $\Sigma$ with all states marked, whether the language $L(A)=\Sigma^*$. 
    To $A$, we construct an NFA $B$ such that $L(B) = @\#L(A) \cup @\Sigma^* \cup \#\Sigma^*$. It is not difficult to construct $B$ from $A$ in polynomial time by adding a new initial state that goes to the initial state of $A$ under the sequence $@\#$ and that has a self-loop under every event from $\Sigma$ after $@$, and by adding a new state reachable under $\#$ having a self-loop under $\Sigma$. Let the abstraction $Q$ remove $\{@\}$, and the observation $P$ remove $\{\#\}$, that is, $\Sigma_{hi}=\Sigma\cup\{\#\}$ and $\Sigma_o=\Sigma\cup\{@\}$. Then $Q(L(B))=\Sigma^* \cup \#\Sigma^*$. We now show that $L(B)$ is OC if and only if $A$ is universal.
  
    If $A$ is universal, then any two different strings $t,t'\in Q(L(B))$ with $P_{hi}(t) = P_{hi}(t')$ are such that $t'=\#t$ (or vice versa). Then, $s=@t$ and $s'=@\#t$ belong to $L(B)$, $Q(s)=t$, $Q(s')=\#t$, and $P(s)=@t=P(s')$. Hence $L(B)$ is OC.
     
    If $A$ is not universal, there is $w\notin L(A)$. Consider the strings $@w$, $\#w \in L(B)$. Then $w$, $\#w\in Q(L(B))$ and $P_{hi}(w)=P_{hi}(\#w)=w$. We now show that there are no strings $s$, $s'\in L(B)$ such that $Q(s)=w$, $Q(s')=\#w$, and $P(s)=P(s')$, i.e., that $L(B)$ is not OC. To do this, we observe that $Q^{-1}(w)\cap L(B) = \{@w\}$ and $Q^{-1}(\#w) \cap L(B) = \{\#w\}$; $@\#w$ does not belong to $L(B)$ because $w\notin L(A)$. But then $P(@w)=@w \neq w = P(\#w)$, which completes the proof.

\subsection{\PSpace-hardness proof of Theorem~\ref{thm7}}
    To show membership in \PSpace, we use a similar technique as in the previous theorem. Namely, we construct an automaton recognizing the sublanguage of $L\times Q(L) \times L \times Q(L)$, where every $(w_1,w_2,w_3,w_4)\in L\times Q(L) \times L \times Q(L)$ satisfies
    \[
      P(w_1)=P(w_3), Q(w_1)=w_2, \text{ and }Q(w_3)=w_4\,.
    \]
    We denote the language by $[L,Q(L),L,Q(L)]$.
    If $L$ is recognized by an NFA $G=(Q,\Sigma,\gamma,I,F)$, then $[L,Q(L),L,Q(L)]$ is recognized by the automaton
    \[
      H=(Q^4, [(\Sigma\cup\{\eps\})\times(\Sigma_{hi}\cup\{\eps\})]^2, \delta, I^4, F^4)
    \]
    where the alphabet consists of quadruples and the transition function $\delta\colon Q^4 \times ((\Sigma\cup\{\eps\})\times (\Sigma_{hi}\cup\{\eps\}))^2 \to Q^4$ is defined on these quadruples as follows:
    \begin{description}
      \item[-] for $a\in\Sigma_o\cap\Sigma_{hi}$, 
        $\delta((p,q,r,s),(a,a,a,a)) = \gamma(p,a) \times \gamma(q,a) \times \gamma(r,q)\times \gamma(s,a)$;
      \item[-] for $a\in\Sigma_o\setminus\Sigma_{hi}$, $\delta((p,q,r,s),(a,\eps,a,\eps)) = \gamma(p,a) \times (\gamma(q,a)\cup\{q\}) \times \gamma(r,a) \times (\gamma(s,a)\cup\{s\})$;
      \item[-] for $a\in\Sigma_{hi}\setminus\Sigma_{o}$, 
        \begin{align*}
        \delta((p,q,r,s),(a,a,\eps,\eps)) & = \gamma(p,a) \times \gamma(q,a)\times \{r\} \times \{s\}, \\
        \delta((p,q,r,s),(\eps,\eps,a,a)) & = \{p\} \times \{q\} \times \gamma(r,a) \times \gamma(s,a);
        \end{align*}
      \item[-] for $a\notin\Sigma_{hi}\cup\Sigma_{o}$,
      \begin{align*}
        \delta((p,q,r,s),(a,\eps,\eps,\eps)) & = \gamma(p,a) \times (\gamma(q,a)\cup\{q\}) \times \{r\} \times \{s\},\\
        \delta((p,q,r,s),(\eps,\eps,a,\eps)) & = \{p\} \times \{q\} \times \gamma(r,a) \times (\gamma(s,a)\cup\{s\}).
      \end{align*}
    \end{description}
    Thus, any element of the language $[L,Q(L),L,Q(L)]$ is of the form $(s,Q(s),s',Q(s'))$ with $P(s)=P(s')$. On the other hand, for any $s,s'\in L$ with $P(s)=P(s')$, it can be shown that $(s,Q(s),s',Q(s'))\in [L,Q(L),L,Q(L)]$. 
    
    Let $e\in\Sigma_c\cap\Sigma_{hi}$. The LOC condition states that for any $s,s'\in L$ with $P(s)=P(s')$ and $Q(s)e,Q(s')e \in Q(L)$ something holds. Therefore, we need to restrict the language $[L,Q(L),L,Q(L)]$ only to the elements for which $Q(s)e,Q(s')e \in Q(L)$. To do this, we concatenate the event $(\eps,e,\eps,e)$ to $[L,Q(L),L,Q(L)]$ and intersect the result with the language $\Sigma^*\times Q(L)\times \Sigma^*\times Q(L)$. This checks that, for any $s,s'\in L$ with $P(s)=P(s')$, we also have $Q(s)e,Q(s')e \in Q(L)$. 
    
    For every such $(s,(Q(s),s',Q(s'))$, the LOC condition requires that there are $u,u'\in (\Sigma\setminus\Sigma_{hi})^*$ such that $P(u)=P(u')$ and $sue,s'u'e\in L$. To verify whether this is satisfied, we check whether the language 
    \[
      [L,Q(L),L,Q(L)] \cdot (\eps,e,\eps,e) \cap (\Sigma^*\times Q(L)\times \Sigma^*\times Q(L))
    \]
    is a subset of the sublanguage $L\times \Sigma^* \times L \times \Sigma^*$ where, for every $(x,y,z,w)$, there is an extension from the language 
    \[
      [(\Sigma\setminus\Sigma_{hi})^*,\eps,(\Sigma\setminus\Sigma_{hi})^*,\eps] \cdot (e,\eps,e,\eps);
    \]
    here, $[(\Sigma\setminus\Sigma_{hi})^*,\eps,(\Sigma\setminus\Sigma_{hi})^*,\eps]$ is the sublanguage of $(\Sigma\setminus\Sigma_{hi})^* \times \{\eps\} \times (\Sigma\setminus\Sigma_{hi})^* \times \{\eps\}$ with the property that, for any $(u,\eps,u',\eps)\in [(\Sigma\setminus\Sigma_{hi})^*,\eps,(\Sigma\setminus\Sigma_{hi})^*,\eps]$, $P(u)=P(u')$. An automaton for this language is constructed in a similar way as the automaton $H$ above. 
    
    Checking the existence of such an extension corresponds to the operation of {\em right quotient\/} denoted by $/$, i.e., we use the language $L\times \Sigma^* \times L \times \Sigma^* / [(\Sigma\setminus\Sigma_{hi})^*,\eps,(\Sigma\setminus\Sigma_{hi})^*,\eps] \cdot (e,\eps,e,\eps)$.\footnote{An automaton for this language can be constructed, e.g., as suggested by Jan Hendrik at https://cs.stackexchange.com/questions/102037/constructive-proof-to-show-the-quotient-of-two-regular-languages-is-regular}
      
    Altogether, for every event $e\in\Sigma_c\cap\Sigma_{hi}$, we check the inclusion
    \begin{align*}
     & [L,Q(L),L,Q(L)] \cdot (\eps,e,\eps,e) \cap (\Sigma^*\times Q(L)\times \Sigma^*\times Q(L)) \\
     \subseteq\ &
      (L\times \Sigma^* \times L \times \Sigma^*)
      /
      [(\Sigma\setminus\Sigma_{hi})^*,\eps,(\Sigma\setminus\Sigma_{hi})^*,\eps]
      \cdot (e,\eps,e,\eps)\,
    \end{align*}
    which requires only polynomial space. We leave the proof details for the full version of the paper.

\begin{figure*}
    \centering
    \includegraphics[scale=.8]{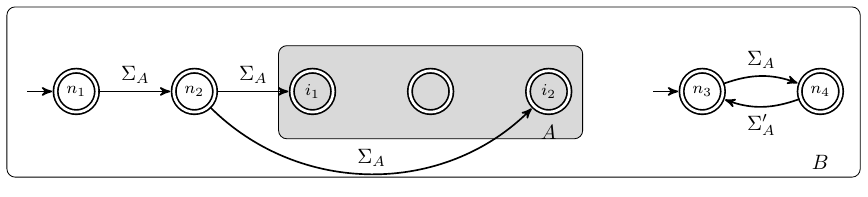}
    \caption{Construction of the NFA $B$ from the NFA $A$; four new states $n_1$, $n_2$, $n_3$, $n_4$; states $n_1$ and $n_3$ are the only initial states of $B$; a transition from $n_1$ to $n_2$, and from $n_2$ to every initial state of $A$ (denoted by $i_1$ and $i_2$) under every event from $\Sigma_A$}
    \label{fig4}
  \end{figure*}

    To show \PSpace-hardness, we reduce the \PSpace-complete universality problem for NFAs with all states marked \cite{KaoRS09}. Let $A$ be an NFA over $\Sigma_A=\{a_1,a_2,\ldots,a_n\}$, $n\ge 2$, such that $L(A)\neq\emptyset$. Then $L(A)$ is prefix-closed, and hence $\eps\in L(A)$. The universality problem asks whether the language $L(A)=\Sigma_A^*$. Let $\Sigma_c=\Sigma_o=\Sigma_{hi}=\Sigma_A$, and let $\Sigma_A'=\{a' \mid a\in\Sigma_A\}$ be a disjoint copy of $\Sigma_A$. From $A$, we construct an NFA $B$ over $\Sigma=\Sigma_A\cup\Sigma_A'$ with all states marked such that 
    $
      L(B) = \overline{\Sigma_A\cdot\Sigma_A\cdot L(A)} \cup \overline{(\Sigma_A\cdot \Sigma_A')^*};
    $
    see Fig.~\ref{fig4} for an illustration. Then, $Q(L(B))=\Sigma_{A}^*$. We now show that $A$ is universal if and only if $L(B)$ is LOC.

    Assume that $A$ is not universal, and consider a shortest string $w\in\Sigma_A^*\setminus L(A)$. Then $w=te$ for some $t\in L(A)$ and $e\in\Sigma_A$. We show that $L(B)$ is not LOC. Set $s=a_1a_1t\in L(B)$ and notice that $e\in \Sigma_A=\Sigma_{hi}\cap\Sigma_c$. Let $t=b_1\cdots b_m$ and $s'=a_1a_1'a_1a_1' b_1b_1'\cdots b_mb_m'e \in L(B)$; indeed, $s'$ can be generated from state $n_3$. Then, $Q(s')e=Q(a_1a_1'a_1a_1' b_1b_1'\cdots b_mb_m')e=Q(s)e\in Q(L(B))$ and $P(s)=s=P(s')$. Since $s$ is generated by $B$ only from state $n_1$, because of the initial prefix $a_1a_1$, and there is no transition labeled by an event from $\Sigma\setminus\Sigma_{hi}=\Sigma_A'$ reachable from $n_1$, there is no $u\in(\Sigma\setminus\Sigma_{hi})^*$ such that $a_1a_1tue=sue \in L(B)$; notice also that $se=a_1a_1te=a_1a_1w\notin L(B)$ because $w\notin L(A)$. Hence, $L(B)$ is not LOC.

    On the other hand, assume that $A$ is universal. Let $s,s'\in L(B)$ be such that $P(s)=P(s')$, and let $e\in\Sigma_c\cap\Sigma_{hi}=\Sigma_A$. Clearly, $Q(s)e,Q(s')e\in Q(L(B))$. Let $t\in\{s,s'\}$. If $t$ is generated from state $n_3$, it can indeed be extended by a string $v\in \Sigma_A'\cup\{\eps\}$ to generate event $e$; in that case, we have that $P(v)=\eps$. If $t$ is generated from state $n_1$, it can clearly generate event $e$ from states $n_1$ and $n_2$; thus, if $t=a_ia_jt'$ for some $a_i,a_j\in\Sigma_A$ and $t'\in\Sigma_A^*$, the universality of $A$ implies that $t'e\in L(A)$. Altogether, we have shown that $sue,s'u'e\in L(B)$ for some $u,u'\in\Sigma_A'^*$ with $P(u)=P(u')=\eps$.

\subsection{Proof of Theorem~\ref{MOCcomplexity}}
    Since MOC is a modification of OC, the proof is a modification of that of Theorem~\ref{thm5}. 
    Let $L \subseteq \Sigma^*$ be a prefix-closed language, and let $\Sigma_o$ and $\Sigma_{hi}$ be the respective observation and high-level alphabets. We show that $L$ is MOC wrt $Q$, $P$, and $P_{hi}$ iff
    \[
      L\parallel_{\Sigma_{hi}\cap \Sigma_o} Q(L) \subseteq Q_2\left(L\parallel_{\Sigma_o} L\right)\,,
    \]
    where, for an event $(a,b)$, $Q_2(a,b)=(a,Q(b))$. Membership in \PSpace then follows, since we can express $Q(L)$, as well as $Q_2(L\parallel_{\Sigma_o} L)$, as NFAs, and the inclusion of two NFAs can be verified in \PSpace.
    
    We first show that if $L$ is MOC, then the inclusion holds. To this end, assume that $(s,t')\in L \parallel_{\Sigma_{hi}\cap\Sigma_o} Q(L)$. By the definition of $\parallel_{\Sigma_{hi}\cap\Sigma_o}$, $s\in L$, $t' \in Q(L)$, and $P_{hi}(Q(s)) = P_{hi}(t')$. Since $L$ is MOC, there is $s' \in L$ such that $Q(s') = t'$ and $P(s) = P(s')$. However, $P(s)=P(s')$ implies $(s,s')\in L\parallel_{\Sigma_o} L$, and $Q(s')=t'$ implies that $(s,Q(s'))=(s,t')$, which shows the inclusion.
    
    We now show that the inclusion implies that $L$ is MOC. To this end, assume that $s\in L$, $t'\in Q(L)$, and $P_{hi}(Q(s))=P_{hi}(t')$. By the definition of $\parallel_{\Sigma_{hi}\cap\Sigma_o}$, we obtain that $(s,t')\in L \parallel_{\Sigma_{hi}\cap\Sigma_o} Q(L)$. Since the inclusion holds, $(s,t')\in Q_2(L \parallel_{\Sigma_o} L)$, which means that there is a pair $(s,s')\in L\parallel_{\Sigma_o} L$ such that $(s,Q(s'))=(s,t')$ and that the strings $s$ and $s'$ belong to $L$ and coincide on $\Sigma_o$, i.e., $P(s)=P(s')$, which was to be shown.
    
    We show \PSpace-hardness by reduction from the problem of deciding universality for NFAs with all states marked. Let $A$ be an NFA over $\Sigma$ with all states marked. 
    We construct a DFA $B$ such that $L(B) = @\#L(A) \cup @\Sigma^* \cup \#\Sigma^* \cup L(A)$. It is not difficult to construct $B$ from $A$ in polynomial time. Let $\Sigma_{hi}=\Sigma\cup\{\#\}$ and $\Sigma_o=\Sigma\cup\{@\}$. Then $Q(L(B))=\Sigma^* \cup \#\Sigma^*$. We now show that $L(B)$ is MOC if and only if $A$ is universal.

    Assume that $L(A)=\Sigma^*$. Let $s\in L(B)$ and $Q(s)\neq t'\in Q(L(B))$ with $P_{hi}(Q(s)) = P_{hi}(t')$. We have the following cases:
    \begin{enumerate}
    \item $Q(s)\in\Sigma^*$ and $t'=\#Q(s)$ for $s\in @\Sigma^* \cup L(A)$:
    \begin{enumerate}
      \item If $s=@w\in @\Sigma^*$, let $s' = @\#w$. 
      \item If $s\in L(A)$, let $s' = \#s$.
    \end{enumerate}
      
    \item $t'\in\Sigma^*$ and $Q(s)=\#t'$ for $s\in @\#L(A) \cup \#\Sigma^*$: 
    \begin{enumerate}
      \item If $s = @\#t' \in @\#L(A)$, let $s' = @t'$.
      \item If $s=\#t'\in\#\Sigma^*$, let $s' = t'$. 
    \end{enumerate}
    \end{enumerate}
    In all cases, it can be verified that $s'\in L(B)$, $Q(s')=t'$, and $P(s)=P(s')$, and hence $L(B)$ is MOC.

    If $A$ is not universal, there is $w\notin L(A)$. We consider the strings $s=@w\in L(B)$ and $\#w\in Q(L(B))$, for which $P_{hi}(Q(@w)) = P_{hi}(\#w) = w$, and show that there is no $s'\in L(B)$ such that $Q(s')=\#w$ and $P(s)=P(s')$, i.e., that $L(B)$ is not MOC. To do this, notice that $Q^{-1}(\#w) \cap L(B) = \{\#w\}$, and hence $P(s) = P(@w) = @w \neq w = P(\#w)$, which completes the proof.

\subsection{Proof of Theorem~\ref{2n}}
  The proof makes use of the following well-known result.
  \begin{lem}[\cite{Won04}]\label{lemma:Wonham}
    Let $\Sigma_{s}\subseteq \Sigma_{hi}$, and let $L_i\subseteq \Sigma_i^*$  be languages, then $Q(\|_{i=1}^n L_i) = \|_{i=1}^n Q_i(L_i) $.
  \end{lem}

    Let $L=\|_{i=1}^n L_i$, and assume that $s\in L$ and $t' \in Q(L)$ are such that $P_{hi}(Q(s)) = P_{hi}(t')$. We show that there is $s''' \in L$ such that $Q(s''') = t'$ and $P(s) = P(s''')$. 
    Since $\Sigma_{s}\subseteq \Sigma_{hi}$, Lemma~\ref{lemma:Wonham} implies $Q(\|_{i=1}^n L_i) = \|_{i=1}^n Q_i(L_i)$. Projecting to local alphabets gives that $P_{i|hi}(Q(s))\in Q_i(L_i)$ and $P_{i|hi}(t')\in Q_i(L_i)$, $i= 1,\dots,n$.
    Moreover, $P_{hi}(Q(s)) = P_{hi}(t')$ implies that $P_{i|hi,o}(P_{hi}(Q(s))) = P_{i|hi,o}(P_{hi}(t'))$. 
    The commutative diagram of Fig.~\ref{fig1} gives that $P_{loc|hi}^{i} P_{i|hi}(Q(s)) = P_{loc|hi}^{i} P_{i|hi}(t')$, and that $P_{i|hi}(Q(s))=Q_iP_i(s)$. 
    Let $s_i=P_i(s)$ and $t_i'=P_{i|hi}(t')$. Then MOC of $L_i$ wrt $Q_i$, $P_{loc}^{i}$, $P_{loc|hi}^{i}$ implies that there is $s_i' \in L_i$ such that $Q_i(s'_i) = t_i'$ and $P_{loc}^{i}(s_i) = P_{loc}^{i}(s'_i)$, $i=1,\ldots,n$. 
    We first show that $\|_{i=1}^n s'_i$ is nonempty. It suffices to prove that  $Q(\|_{i=1}^n s_i')$ is nonempty. Since $\Sigma_{s}\subseteq \Sigma_{hi}$, Lemma~\ref{lemma:Wonham} gives that $Q(\|_{i=1}^n s_i') = \|_{i=1}^n Q_i(s_i') = \|_{i=1}^n P_{i|hi}(t')$, which is nonempty, because $t' \in \|_{i=1}^n P_{i|hi}(t')$. Hence, there is $s'\in \|_{i=1}^{n} s_i'$ such that $Q(s')=t'$. 
    Furthermore, $P_{loc}^{i}(s_i) = P_{loc}^{i}(s'_i)$, for $i=1,\dots,n$, means that $P(s)\in P(\|_{i=1}^{n} s_i) = \|_{i=1}^{n} P^{i}_{loc}(s_i) = \|_{i=1}^n P_{loc}^{i}(s'_i)=P(\|_{i=1}^n {s'_i}) \ni P(s')$ by $\Sigma_{s}\subseteq \Sigma_{o}$ and Lemma~\ref{lemma:Wonham}. Hence, there is $s''\in \|_{i=1}^{n} s_i'$ such that $P(s)=P(s'')$. 
    Since $P_{hi}(t')=P_{hi}Q(s)$, there is $s'''\in Q(s')\|P(s'')$. But then $Q(s''')=Q(s')=t'$ and $P(s''')=P(s'')=P(s)$, which was to be shown.
    
\end{document}